# Reversible Projective Measurement in Quantum Ensembles

Anatoly Khitrin,[1] Michael Michalski,[2] and Jae-Seung Lee[3]

[1]Department of Chemistry, Kent State University, Kent OH 44242-0001

[2]Department of Physics, Kent State University, Kent OH 44242-0001

[3]Department of Chemistry, New York University, New York, NY 10003

*Abstract*

We present experimental NMR demonstration of a scheme of reversible projective measurement, which allows extracting information on outcomes and probabilities of a projective measurement in a non-destructive way, with a minimal net effect on the quantum state of an ensemble. The scheme uses reversible dynamics and weak measurement of the intermediate state. The experimental system is an ensemble of $^{133}$Cs ($S = 7/2$) nuclei in a liquid-crystalline matrix.

**INTRODUCTION**

The result of quantum-mechanical measurement is, in general, probabilistic. It returns one of several possible eigenvalues of the measured quantity and, after this value is obtained, changes (collapses) the state of the system to the corresponding eigenstate [1]. The initial state of the system is irreversibly spoiled by this process. A single measurement does not provide information about the initial state. In order to retrieve this information one needs to perform the measurement many times, starting with the same initial state (full reconstruction requires measurements in different bases). Therefore, an ensemble of measurements is to be considered. Alternatively, one can think about these measurements as being performed simultaneously on an ensemble of systems in the same initial state. Formally, this equivalence is expressed by the fact that the density matrix provides a complete description of a quantum system [2].

Both quantum-mechanical evolution and measurement are linear in the density matrix. This means, in particular, that if some dynamical process produces a small change of the density matrix, it can be equivalently described as resulting from small changes of states of individual systems or from large changes for a small fraction of the systems. Completeness of description, provided by the density matrix, guarantees that no physical experiment can discriminate between these two alternatives. In other words, any interpretation of a quantum-mechanical experiment, consistent with the density-matrix formalism, is a valid interpretation.

The idea of our experiment can be illustrated by a scheme of reversible Stern-Gerlach experiment [3] in Fig. 1. The theory of reversible Stern-Gerlach experiment has been developed in Ref. [4]. A beam of particles in the same spin state is split according to the projection of their spins on the direction of the magnetic field gradient. Without any measurement, the reversed gradients can merge the particles into a single beam, in which all the particles will have the same initial spin state. A weak, almost transparent, detector can use a tiny fraction of the particles to perform projective measurement, while the spin state of the majority of the particles does not change. Such measurement can provide information about possible outcomes of projective measurement and, at the same time, produce negligible effect on the system's density matrix. It can be mentioned that this scheme, using a simple notion that in a big ensemble of systems only a small fraction of systems can be used in projective measurement, is different from the concept of interaction-free measurement, which has been proposed in [5,6] and discussed more recently in [7].

In nuclear magnetic resonance (NMR), the observable is a transverse magnetization of the entire sample, which is weakly coupled to the radio-frequency coil. The measurement is, therefore, a weak ensemble measurement, producing small effect on the system's density matrix. However, NMR experiment can be designed to provide information on possible outcomes of projective measurement and probabilities of such outcomes [8]. It is interesting that there exists flexibility in choosing the basis for the projection and, as an example, the states can be projected on eigenstates of an artificially built average Hamiltonian [9].

In this paper, we describe an NMR experiment, where extraction of information on outcomes of projective measurement is followed by a reverse evolution, returning the system to its initial state. The possibility of getting the results of a projective measurement without spoiling the density matrix would add flexibility to designing new quantum algorithms: quantum evolution can now be conditioned by the results of intermediate measurements, similar to how in classical computation evolution is conditioned by intermediate states of a computer.

## I. EXPERIMENTAL

The sample we used in our experiment is a 50% solution of cesium pentadecafluorooctanoate in $D_2O$, which can form a discotic nematic liquid-crystalline phase [10,11]. The desired phase exists in a narrow temperature interval. Our experiments have been performed at 48ºC. Necessary mesophase has been reached by 30 min constant-rate cooling starting with isotropic state at 60ºC. Due to anisotropic environment, the cesium nuclei (spin $S=7/2$) experience residual quadrupolar coupling. The corresponding spin Hamiltonian has the form

$$H = \omega_0 S_Z + (\omega_q/2) S_Z^2 , \qquad (1)$$

where $\omega_0$ is the Larmor frequency, $\omega_q$ describes the strength of residual quadrupolar coupling, and $S_Z$ is the projection of the cesium nuclear spin on the direction of external magnetic field. The equilibrium $^{133}$Cs NMR spectrum, shown in Fig. 4a, consists of 7 equidistant peaks, separated by $\omega_q$. The peaks correspond to allowed single-quantum transitions ($|\Delta m|=1$, where the magnetic quantum number $m$ is the eigenvalue of $S_Z$) between 8 energy levels. This system is a convenient 3-qubit model with the simplest possible spectrum, which allows addressing individual qubits and measuring all the quantum states. It has been used before for implementing various quantum algorithms [11-14].

A pure quantum state is always an approximation for any real physical system. While for many systems the thermal equilibrium state can well approximate a pure state, for nuclear spins many energy levels are almost equally populated. The idea of a pseudopure state [15,16] is frequently used in NMR implementations to reproduce the pure-state behavior. The density matrix of the pseodopure state is a sum of two terms: one proportional to the identity operator, and another proportional to the density matrix of a pure state. The identity matrix does not evolve nor contribute to observables. Therefore the behavior (results of physical experiments) is exactly the same as for the same system in a true pure state. The simplest way to prepare a pseudopure state is to equalize populations of all but one quantum level, as it is schematically shown in Fig. 2. An efficient method of equalizing populations is described in Ref. [11]. It uses multi-frequency irradiation, with adjusted amplitudes of harmonics, to simultaneously drive necessary populations of levels so that they all reach the same value at the same moment of time. This approach was used in the present work for the pseudopure state preparation.

The NMR experiment, described below, follows the logic of the scheme in Fig. 1. The pulse sequence is shown in Fig. 3. It starts with a six-frequency pulse (A), labeled sh$_{\varphi 1}$, to equalize populations of the upper seven energy levels [11]. The result of this shaped pulse is the creation of the pseudopure ground state with $m = 7/2$.

Part B is a composite 90º-pulse consisting of two 45º-pulses and free evolution time $t_c$ − δ between these pulses, which works as follows. Because all the neighboring peaks in the spectrum are separated by the same frequency $\omega_q$, free evolution with the Hamiltonian (1) is cyclic with the period $t_c = 2\pi/\omega_q$. Evolution during the time $t_c$ − δ is, therefore, equivalent to a time-reversed evolution lasting the time δ. δ is experimentally adjustable parameter, approximately equal to the total duration of the two 45º-pulses, and it is used to compensate the action of the quadrupolar coupling during the action of the radio-frequency (RF) pulses. The reason we replaced a simple 90º-pulse by a composite pulse is that the RF field available in our experiment was insufficiently large compared to $\omega_q$ and, as a result, a single pulse provided poor accuracy of 90º rotation.

Part C is the first signal acquisition after a small-angle reading pulse. A small flip angle of this pulse is needed to avoid mixing of the different levels' populations and to provide a linear-response signal. Fourier transform of this signal gives a linear-response spectrum, where intensities of individual peaks are proportional to differences of populations for the corresponding pairs of energy levels. The length of the part C is chosen to be multiples of the cycle times $t_c$ to avoid any change of the state by a free evolution with the Hamiltonian (1). Step D is the composite 90º-pulse with the direction opposite to that in step B. Step E contains another small-angle reading pulse and the second signal acquisition.

The phases $\varphi 1 - \varphi 4$, shown as subscripts in Fig. 3, have been independently cycled to eliminate unwanted coherences. As an example, the composite 90º-pulse in B (phase $\varphi 2$) performs a rotation from Z- to X-basis. Cycling the relative phase between $\varphi 2$ and the first acquisition phase $\varphi 3$ in C eliminates coherences in X-basis and accomplishes the projection, making, therefore, the signal in C dependent only on differences of populations in X-basis.

All the spectra have been recorded at 48ºC using Varian Unity/Inova 500 MHz NMR spectrometer (11.7 T magnet).

## II. RESULTS AND DISCUSSION

Fig. 4a shows the $^{133}$Cs thermal equilibrium spectrum for our system. Since $\omega_0 \gg \omega_q$, the differences of populations for all allowed transitions are equal. Theoretical intensities are proportional to the differences of populations (equal at equilibrium) and squares of the transition matrix elements for the operator $S_X$. We have found that the relative intensities of peaks in experimental spectrum in Fig. 4a are very close to the theoretical values (i.e. the squares of matrix elements). We have used the experimental equilibrium intensities in Fig. 4a to "normalize" all other spectra and eliminate the dependence on matrix elements as following. Each peak's intensity in a non-equilibrium spectrum has been divided by the intensity of corresponding peak in the equilibrium spectrum to give a value proportional to the difference of populations.

The pseudopure state with $m = 7/2$ has been prepared by applying the shaped six-frequency pulse as described in Ref. [11]. The linear-response spectrum for this state (Fig. 4b) shows one transition from $m = 7/2$ to $m = 5/2$ state. For a pseudopure state, the essential part of the density matrix is not equal but proportional to the density matrix of corresponding pure state. For the single-spin density matrix of pure state with $m = 7/2$ the difference of populations (diagonal matrix elements) between the states with $m = 7/2$ and $m = 5/2$ is one. To set the correct scale of the population differences, we assigned the intensity of peak in Fig. 4b to correspond to the populations difference one. After this adjustment, sum of all populations should be one, and each population can be viewed as probability to be in a given state.

The spectrum in Fig. 4c shows the result of applying steps A-C of the pulse sequence (Fig. 3). Step A prepares a system in eigenstate of $S_Z$ with eigenvalue $m = 7/2$. The composite 90º-pulse in step B performs rotation of the state, which can be also viewed as rotation of the basis, i.e. a transformation from Z-basis to X-basis. The eigenstate in Z-basis is not an eigenstate in X-basis and, after projection on eigenfunctions of $S_X$, the system can be found in any of the eight $S_X$ eigenstates with some probabilities. These probabilities (populations) have been calculated from the spectrum in Fig. 4c by using the "normalization" described above. Populations have been obtained from their differences by integration and setting the integration constant so that the sum of all populations is one.

Experimentally measured populations are shown in Fig. 5 as narrow filled bars. One can notice one negative value for $S_X = -7/2$. This is the result of experimental errors and the fact that we did not explicitly use the restriction that all the probabilities (populations) are positive. Theoretically, the probabilities can be calculated from the corresponding Clebsch-Gordan coefficients [17]. Table 1 lists the eigenfunctions of $S_Z$ in the basis of eigenfunctions of $S_X$. As one can see from this table, for the eigenstate of $S_Z$ with the eigenvalue $m = 7/2$, the probabilities for the spin to have different projections on X-axis (in increasing order of projections) are 1/128, 7/128, 21/128, 35/128, 35/128, 21/128, 7/128, and 1/128. These theoretical probabilities are shown in Fig. 5 as wide bars. One can see reasonable agreement between the experimental and theoretical values.

The spectrum in Fig. 4c has been recorded by using sufficiently long acquisition time, in order to make it look "normal" and free of distortions. The spectrum is equidistant, and the only information we need are seven complex numbers, the amplitudes of the peaks. They can be extracted as Fourier coefficients from a signal acquired within a single period $t_c$. Longer acquisition times (which should be multiples of $t_c$) make it more difficult to reverse the evolution and to return the system to its initial state. We have found that at $(4 \sim 6)\ t_c$ the reversibility is still reasonably good. At the same time, such acquisition times make it possible to Fourier-transform the signals, using standard spectrometer software and large broadening factor, to make the result look like spectrum. The result of the reversal in Fig. 4d shows Fourier transform of the second signal when the duration of the first acquisition was $4\ t_c$. Again, the length of the second acquisition is made sufficiently long. For ideal reversal, one would expect the spectrum in Fig. 4d to coincide with the spectrum in Fig. 4b. One can notice some degradation of the peak intensity and appearance of small unwanted peaks in the central part of the spectrum. They are the results of experimental errors and relaxation, contributing to non-perfect reversibility. When comparing the intensities of the peaks in the central part with the intensity of the main peak in Fig 4c, one should keep in mind that originally the intensities of central peaks are larger due to larger transition matrix elements.

Finally, Fig. 6 shows the spectra when two signals are acquired within the same scan with the same short acquisition times $4\ t_c$. The spectrum in Fig. 6a shows the result of projection (cf. Fig. 4c), and the spectrum in Fig. 6b shows the recovered pseudopure ground state (cf. Fig. 4d). Of course, both spectra are severely distorted due to very short acquisition times, but these distortions are consistent with the expected result of the signals truncation and large broadening factor used in Fourier transform.

It should be mentioned that all the spectra presented above have been recorded in the same experimental run, using the same experimental parameters: pulses durations and power, delays, and phasing of the spectra (with the exception of different acquisition times in Figs. 4 and 6). Due to large number of adjustable experimental parameters, each individual spectrum could be made looking almost ideal by using individual tuning of the parameters and using more transients for better signal-to-noise. However, we decided that picking the best individual spectra would not be a fair way of presenting the experimental data. The reason is that the liquid-crystalline phase we used is not very stable. Separation between peaks in the spectrum slowly changes with time, and each time the phase is newly prepared, the separation is slightly different. This needs an adjustment of the parameters of the six-frequency pulse and inter-pulse delays in the pulse sequence. The requirement of keeping all the parameters unchanged restricted the experimental time.

### III. CONCLUSIONS

The projective measurement in our experiment is performed not by irreversible destruction of coherences but by temporal averaging. The same experiment is repeated many times at different phases of the RF pulses and the spectra acquired in individual scans are added together. One can imagine similar experiments performed in parallel. The ensemble of systems (nuclear spins) can be divided into sub-ensembles (64 in our case). The experiments with different phases are performed simultaneously for all sub-ensembles, and the results of all measurements are added. Each of the sub-ensembles follows a reversible dynamics and, in the end, returns to its initial state. As a consequence, the total ensemble will be returned to its initial state, while the intermediate result, obtained by summation of the weak-measurement results for individual sub-ensembles, provides information on possible outcomes of projective measurement and the probabilities of such outcomes. Therefore, for large ensembles of quantum systems, it is possible to design quantum algorithms where evolution is conditioned by intermediate results of projective measurements, performed in a non-destructive way.

**Tables**

Table 1. Eigenvectors of $S_Z$ with different eigenvalues $m$ for spin 7/2 in the $S_X$ basis.

| $m$ | Eigenvectors of $S_Z$ in the $S_X$ basis |
|---|---|
| -7/2 | $\frac{1}{16}\left(\sqrt{2}, \sqrt{14}, -\sqrt{42}, \sqrt{70}, -\sqrt{70}, \sqrt{42}, -\sqrt{14}, \sqrt{2}\right)$ |
| -5/2 | $\frac{1}{16}\left(\sqrt{14}, -5\sqrt{2}, 3\sqrt{6}, -\sqrt{10}, -\sqrt{10}, 3\sqrt{6}, -5\sqrt{2}, \sqrt{14}\right)$ |
| -3/2 | $\frac{1}{16}\left(-\sqrt{42}, 3\sqrt{6}, -\sqrt{2}, -\sqrt{30}, \sqrt{30}, \sqrt{2}, -3\sqrt{6}, \sqrt{42}\right)$ |
| -1/2 | $\frac{1}{16}\left(\sqrt{70}, -\sqrt{10}, -\sqrt{30}, 3\sqrt{2}, 3\sqrt{2}, -\sqrt{30}, -\sqrt{10}, \sqrt{70}\right)$ |
| 1/2 | $\frac{1}{16}\left(-\sqrt{70}, -\sqrt{10}, \sqrt{30}, 3\sqrt{2}, -3\sqrt{2}, -\sqrt{30}, \sqrt{10}, \sqrt{70}\right)$ |
| 3/2 | $\frac{1}{16}\left(\sqrt{42}, 3\sqrt{6}, \sqrt{2}, -\sqrt{30}, -\sqrt{30}, \sqrt{2}, 3\sqrt{6}, \sqrt{42}\right)$ |
| 5/2 | $\frac{1}{16}\left(-\sqrt{14}, -5\sqrt{2}, -3\sqrt{6}, -\sqrt{10}, \sqrt{10}, 3\sqrt{6}, 5\sqrt{2}, \sqrt{14}\right)$ |
| 7/2 | $\frac{1}{16}\left(\sqrt{2}, \sqrt{14}, \sqrt{42}, \sqrt{70}, \sqrt{70}, \sqrt{42}, \sqrt{14}, \sqrt{2}\right)$ |

**Figure captions**

FIG. 1. Scheme of reversible Stern-Gerlach experiment with weak measurement.

FIG. 2. Scheme of the levels populations and observed transitions in linear-response spectra for a) thermal equilibrium and b) pseudopure ground state.

FIG. 3. Pulse sequence: (Step A) a shaped pulse creating pseudopure state; (Step B) a composite 90º pulse; (Step C) a small-angle reading pulse and first signal acquisition; (Step D) a reversed composite 90º pulse; (Step E) a small-angle reading pulse and second signal acquisition.

FIG. 4. $^{133}$Cs linear-response NMR spectra for (a) thermal equilibrium state, (b) pseudopure state with $S_Z=7/2$, (c) projection of this state on $S_X$ eigenfunctions, and (d) recovered pseudopure state with $S_Z=7/2$ .

FIG. 5. Populations of states after projections (wide bars are theoretical values).

FIG. 6. Spectra obtained from two short acquisitions of $4t_c$ duration within the same experiment: a) first acquisition (cf. Fig. 4c); b) second acquisition (cf. Fig. 4d).

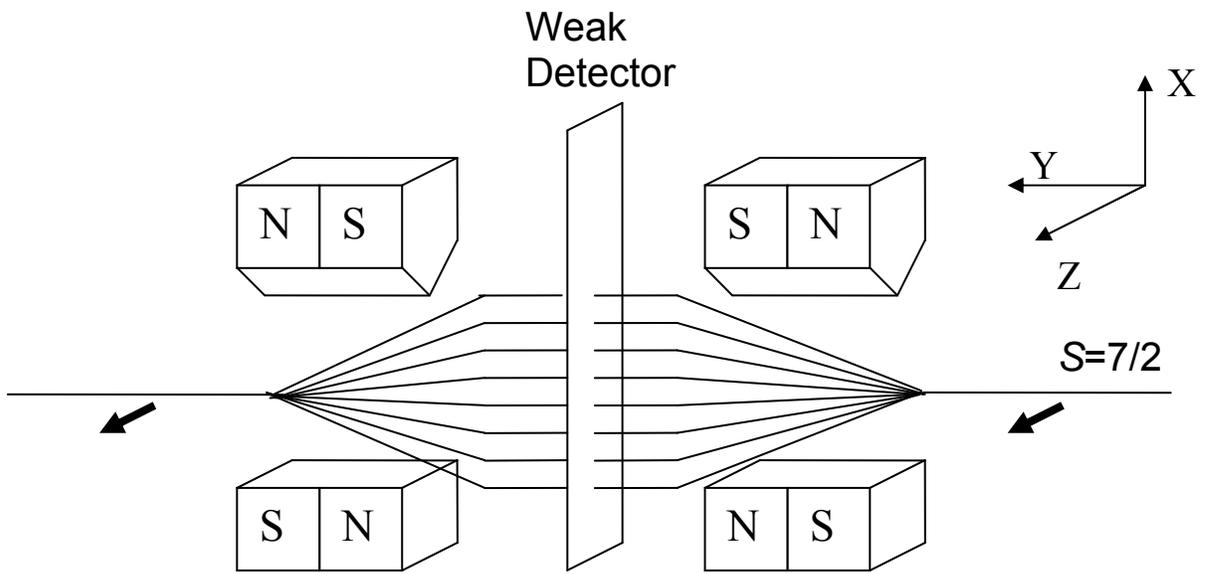

**Fig. 1**

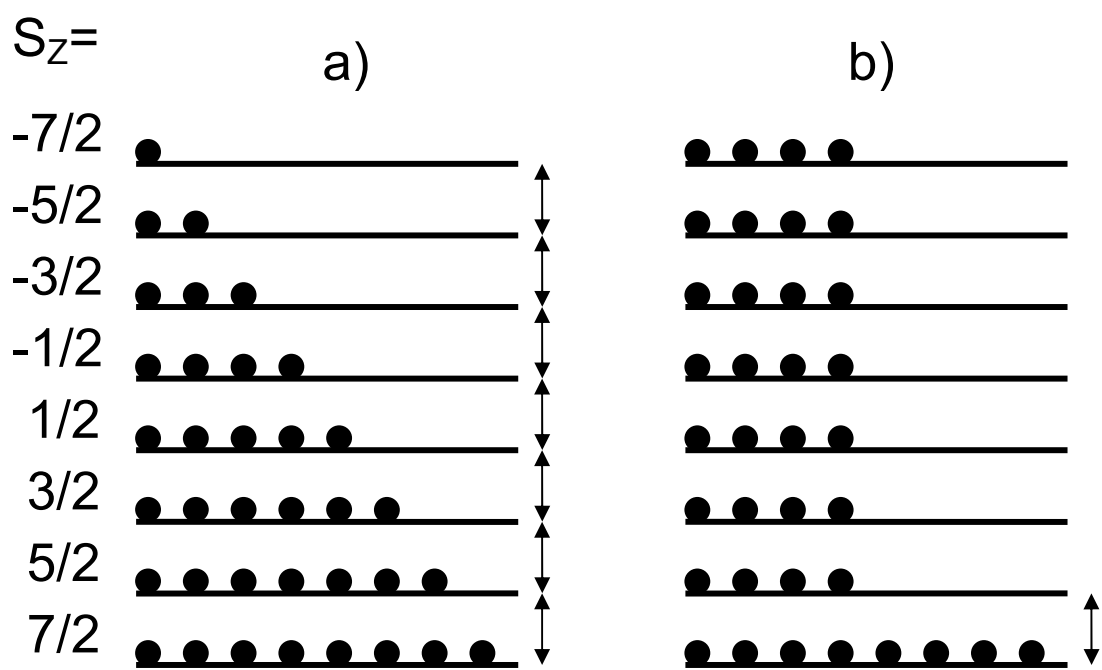

**Fig. 2**

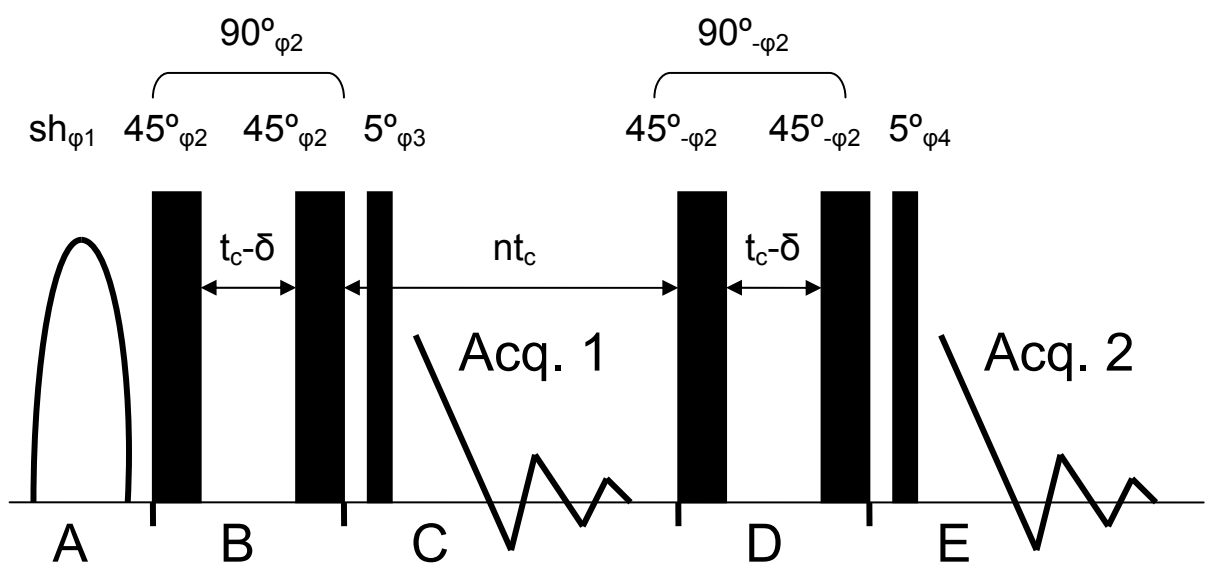

**Fig. 3**

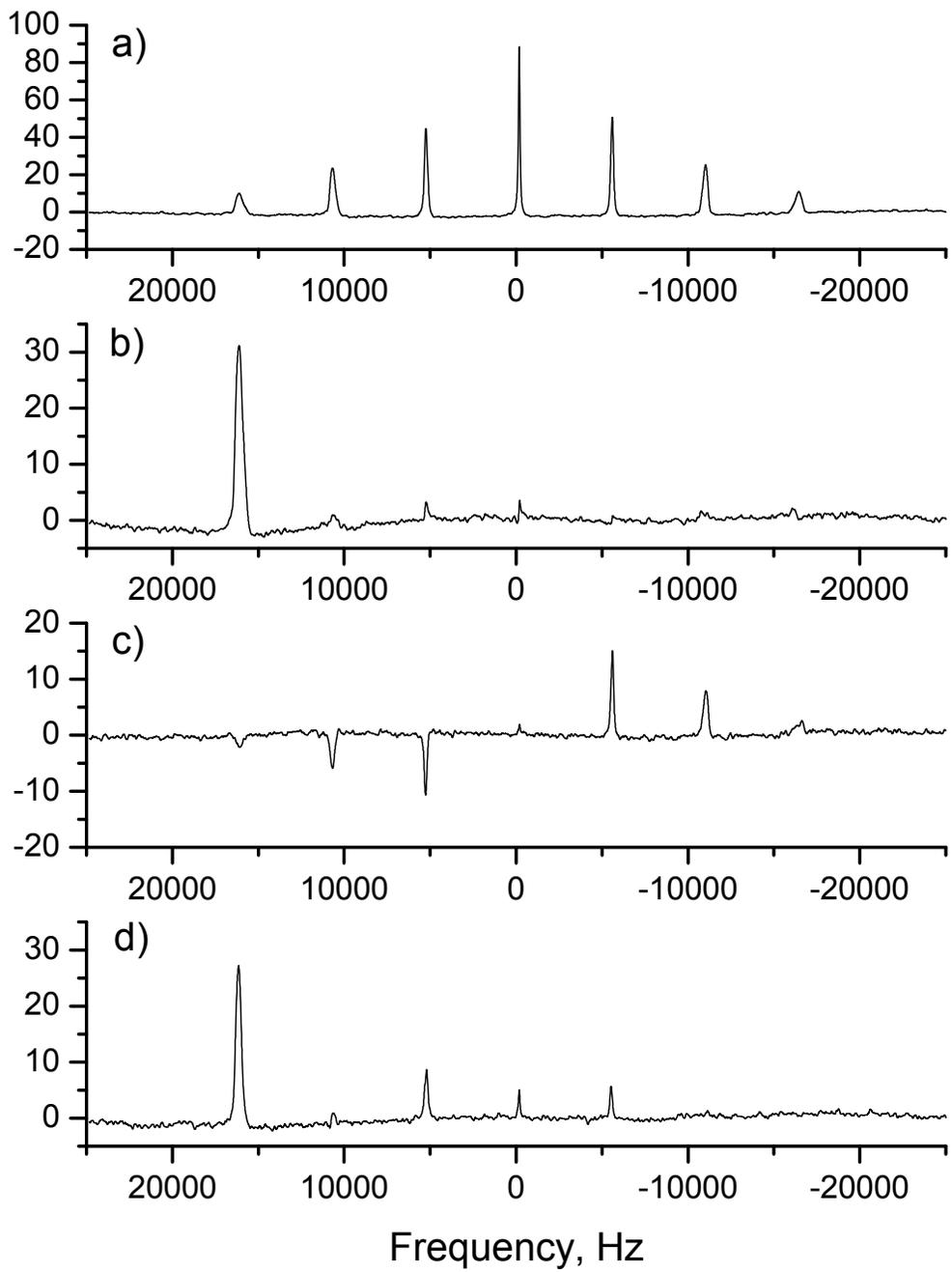

**Fig. 4**

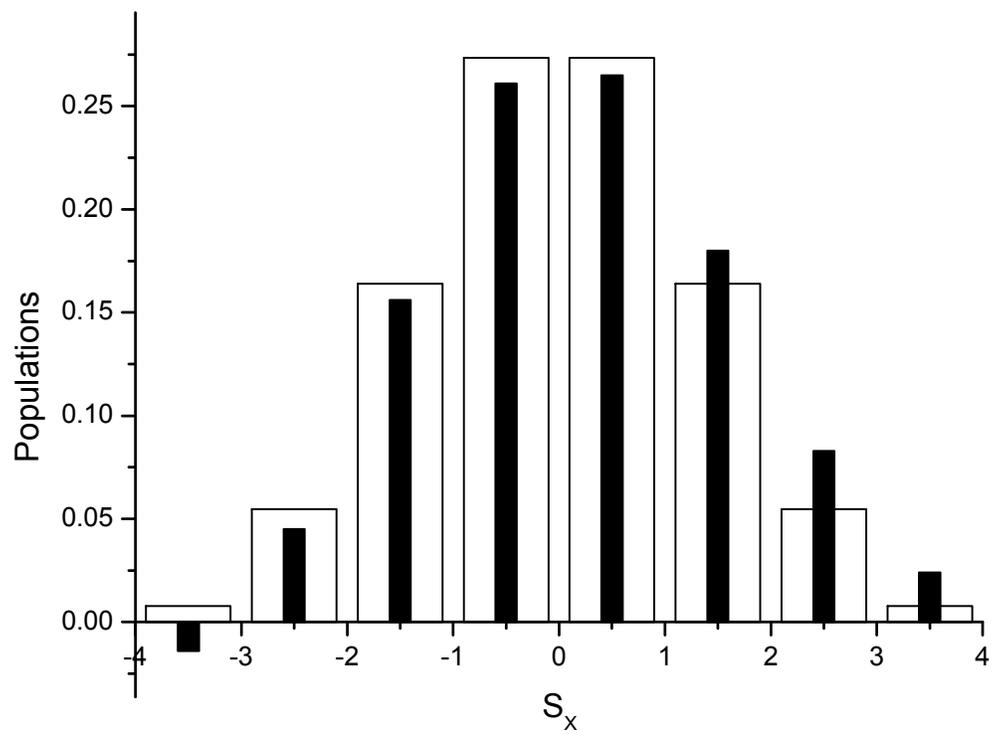

**Fig.5**

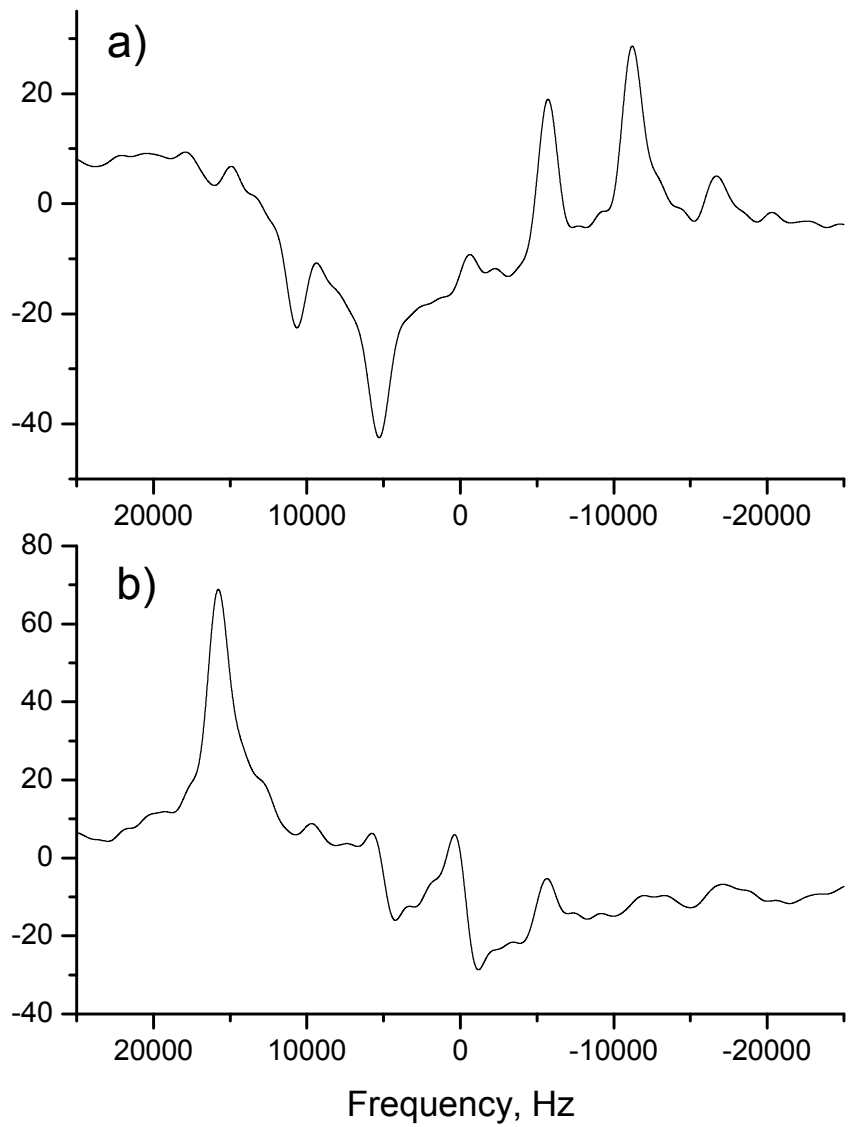

**Fig. 6**